\documentclass[journal=nalefd,manuscript=letter]{achemso}

\usepackage{graphicx}
\usepackage{dcolumn}
\usepackage{bm}
\usepackage{mathptmx}
\usepackage{braket}
\usepackage{amsmath}
\usepackage{hyperref}
\hypersetup{
    colorlinks=true,
    linkcolor=blue,     
    urlcolor=blue,
    citecolor=blue,
}
\usepackage[capitalise]{cleveref}
\usepackage{xcolor}
\usepackage[version=4]{mhchem}
\usepackage{float}
\usepackage[mathlines]{lineno}
\usepackage{comment}
\usepackage{mathtools}
\setkeys{acs}{etalmode=truncate,maxauthors=0}

\DeclarePairedDelimiter\abs{\lvert}{\rvert}
\DeclarePairedDelimiter\norm{\lVert}{\rVert}

\author{Kin On Ho}
\author{Hamza Abudayyeh}
\author{Andrew Perez}
\affiliation
{Department of Physics, The University of Texas at Austin, Austin, TX, USA}
\author{Hui-Ping Chang}
\affiliation
{Department of Electrical and Computer Engineering, The University of Texas at Austin, Austin, TX, USA}
\alsoaffiliation
{Microelectronics Research Center, The University of Texas at Austin, Austin, TX, USA}
\author{Liang Juan Chang}
\affiliation
{Institute of Space Systems Engineering, National Yang-Ming Chiao-Tung University, Hsinchu 30010, Taiwan}
\author{Junwei Tong}
\affiliation
{Department of Physics, The University of Texas at Austin, Austin, TX, USA}
\author{Can Cui}
\affiliation
{Department of Electrical and Computer Engineering, The University of Texas at Austin, Austin, TX, USA}
\alsoaffiliation
{Microelectronics Research Center, The University of Texas at Austin, Austin, TX, USA}
\author{Jean Anne C. Incorvia}
\affiliation
{Department of Electrical and Computer Engineering, The University of Texas at Austin, Austin, TX, USA}
\alsoaffiliation
{Microelectronics Research Center, The University of Texas at Austin, Austin, TX, USA}
\email{incorvia@austin.utexas.edu}
\author{Xiaoqin Li}
\affiliation
{Department of Physics, The University of Texas at Austin, Austin, TX, USA}
\email{elaineli@physics.utexas.edu}

\title[Nanodiamond Sensing]
    {Nanodiamond Sensing of Stray Fields during Domain Reversal}


\begin{document}


\begin{abstract}
Quantitative measurement of nanoscale stray magnetic fields during domain reversal is important for understanding and optimizing magnetic memory and logic devices. Yet, achieving high spatial resolution with minimally invasive probes remains challenging. Here we demonstrate nanodiamonds (NDs) hosting an ensemble nitrogen vacancy (NV) centers as local quantum sensors of domain-reversal stray fields in a CoFeB strip. During magnetic-field-driven reversal, the ND sensors resolve three distinct local responses depending on their positions relative to the strip: a positive frequency jump at the edge, an unexpected negative jump just outside the strip, and a negligible change near the center. These contrasting signals are quantitatively explained by magnetostatic boundary fields from micromagnetic simulations. We further implement a fully constrained fitting procedure for optically detected magnetic resonance spectra, enabling robust field extraction even when resonances are partially resolved. Together, these results establish NV-ND magnetometry as a simple, transferable, and quantitative approach for characterizing spatially heterogeneous magnetic fields in thin-films and spintronic devices.
\end{abstract}

Magnetic stray fields produced by domain structures encode key information about magnetization reversal, domain wall motion, and device operation in nanoscale magnets. Quantitative access to these local fields is increasingly important for thin-film spintronic platforms, including magnetic random-access memory (MRAM), domain wall devices, magnetic material based computing, and related architectures in which magnetic textures directly govern functionality. Yet these stray fields are inherently nonuniform and short-ranged, making them challenging to measure with both nanoscale spatial sensitivity and minimal sample perturbation. Existing magnetic characterization techniques provide only partial solutions. Bulk probes, such as vibrating sample magnetometry and superconducting quantum interference device (SQUID) magnetometry, lack spatial resolution. Conventional magneto-optical Kerr effect (MOKE) measurements are typically limited to micron-scale resolution and are not intrinsically quantitative. Magnetic force microscopy can achieve high spatial resolution, but tip-induced perturbations often complicate the interpretation of the native magnetic texture. Consequently, there remains a need for experimentally accessible methods that can quantitatively probe local magnetic fields during switching in patterned magnetic devices. 

Over the past decade, nitrogen vacancy (NV) centers in diamond have emerged as a pioneer magnetometer for studying magnetic materials~\cite{Rondin2014Magnetometry, Casola2018Probing, Scholten2021Widefield}. Techniques such as single NV scanning probe microscopy (NV-SPM)~\cite{Rondin2013Stray, Tetienne2015The, Gross2016Direct, Yu2018Room, Jenkins2019Single, Tsukamoto2025Observation} and ensemble NV widefield imaging (NV-WFM)~\cite{Bindu2025Quantum, Broadway2020Imaging, Healey2022Varied, Huang2023Revealing, Robertson2023Imaging, Yan2022Quantum, Kitagawa2024Wide} have enabled detailed visualization of magnetic textures, including domain structures and domain wall types. For example, by probing stray magnetic fields in close proximity ($\approx120$~nm) to ultrathin samples, NV-SPM was applied to distinguish Bloch- and Néel-type domain walls~\cite{Tetienne2015The}. However, these prior studies have predominantly probed static magnetic configurations, providing only snapshots of the magnetic state. In contrast, many functional magnetic devices operate through field-driven evolution, where domain reversal under a sweeping magnetic field is central to device functionality. Direct and spatially resolved measurements of such dynamic processes remain scarce, in part due to the short-range and highly nonuniform nature of stray magnetic fields during domain evolution. As an intermediate approach bridging NV-SPM and NV-WFM, nanodiamond (ND)-based sensing offers a compelling combination of nanoscale proximity and experimental simplicity. NDs can be deployed directly onto devices without the need for specialized scanning probe instrumentation, while their small size (down to $\approx 30$~nm) enables spatial resolution beyond the optical diffraction limit. Despite these advantages, the huge potential of ND-based sensing for probing field-driven magnetic dynamics remains largely unexplored, with only limited prior work on domain wall oscillations~\cite{Rable2023Off}.

In this work, we demonstrate the application of ND-based magnetometry for probing magnetic domain flips beyond the optical diffraction limit. We perform quantum sensing using 140-nm NDs on a ferromagnetic CoFeB thin film strip, a system previously studied using NV techniques due to its rich domain physics~\cite{Tetienne2015The, Gross2016Direct, Kitagawa2024Wide}, frequency conversion~\cite{Wu2024Wideband}, and room-temperature skyrmions~\cite{Yu2018Room, Jenkins2019Single}. Leveraging the perpendicular magnetic anisotropy of CoFeB, we investigate domain switching during magnetic field sweeps. We observe both expected (increasing) magnetic field jumps and unexpected (decreasing) jumps, which are corroborated by independent micromagnetic simulations. We analyze the ODMR spectra using a fully constrained NV fitting procedure that incorporates the NV Hamiltonian and the full spectral lineshape. Compared with conventional multi-peak analysis, this approach makes use of the full ODMR information and allows reliable reconstruction of the local magnetic field even when only part of the spectrum is experimentally accessible. Finally, we discuss the practical advantages of ND-based sensing compared to other techniques. Together, these results establish NDs as a powerful and versatile quantum-sensing platform for quantitative characterization of spatially heterogeneous magnetic fields in magnetic devices and materials.

\section{Results and Discussion}
\subsection{Device configuration}
\begin{figure}[t]
    \includegraphics[width=12cm]{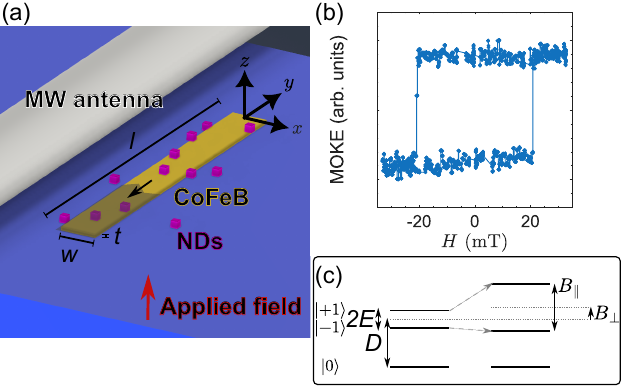}
    \caption{Experimental principle. (a) Illustration of the experimental setup. The CoFeB strip fabricated on a Si substrate has approximate dimensions $(\mathrm{width}~w)~5~\mathrm{\mu m}~\times~(\mathrm{length}~l)~83~\mathrm{\mu m}~\times~(\mathrm{thickness}~t)~1~\mathrm{nm}$. 140-nm NDs serving as local magnetic field sensors, and a Pt wire acts as a MW antenna. An external magnetic field applied perpendicular to the film drives domain flip, modifying the stray field measured by the ND sensors. (b) MOKE measurement of the CoFeB strip, yielding a coercive field $H_{c} = 21.0\pm0.4$~mT. (c) Simplified NV energy diagram depicting zero-field splitting parameters $D$ and $E$, and the energy shift from the Zeeman effect due to the B field parallel and perpendicular to the NV axis.}
    \label{fig:fig1}
\end{figure}

\cref{fig:fig1}(a) illustrates the experimental setup. 140-nm NDs are randomly distributed around the CoFeB strip. A Pt wire serves as a microwave (MW) antenna for optically detected magnetic resonance (ODMR) measurements~\cite{Gruber1997Scanning}. The hysteresis loop as a function of the applied magnetic field $H$ perpendicular to the strip is shown in \cref{fig:fig1}(b), measured using a standard polar MOKE microscopy setup. The MOKE signal is sensitive to the magnetization perpendicular to the CoFeB film, but it cannot quantitatively determine the magnetization or the stray magnetic field. From this measurement, we extract a coercive field of $H_{c} = 21.0\pm0.4$~mT, corresponding to the critical field for domain reversal. Such domain flips modify the local stray field and thereby alter the NV energy levels. As illustrated in \cref{fig:fig1}(c), a longitudinal magnetic field $B_{\parallel}$ splits the NV spin states, whereas a transverse magnetic field $B_{\perp}$ shifts the NV resonance frequencies to higher values.

\subsection{Fully constrained NV fitting}
Conventional analysis of ODMR spectra typically relies on multi-peak fitting (most commonly using Lorentzian functions) followed by analyzing the magnetic field with analytical expressions. While widely used, this approach has two important limitations in the present experiment. First, only the resonance frequencies are used in the field calculation, while other lineshape parameters (e.g., contrast and linewidth) are discarded, resulting in a loss of information encoded in the ODMR spectra. Secondly, depending on the analysis framework -- such as perturbation theory~\cite{Barry2020Sensitivity}, geometric approaches~\cite{Zhang2022Single, Maertz2010Vector}, or cone-optimization~\cite{Yip2019Measuring, Weggler2020Determination} -- three NV orientations (i.e., six ODMR resonances) are typically required to uniquely determine the magnetic field vector, which in practice demands broadband MW excitation to access the $\ket{0} \rightarrow \ket{+1}$ transitions at higher frequencies under relatively large magnetic fields. In our measurements, this requirement becomes increasingly restrictive in the high-field regime, where some resonances fall outside the accessible frequency windows ($>15~\rm{mT}$ in our case).

To overcome these limitations, we adopt a different strategy by performing a fully constrained NV fitting procedure that models the entire ODMR spectrum rather than only the fitted peak positions. The method combines the exact NV spin Hamiltonian with a phenomenological lineshape model, and uses the full spectral response to reconstruct the local magnetic field. This approach is motivated by an earlier work on NV-based magnetic-field reconstruction in superconductors~\cite{Ho2026Widefield} and is particularly advantageous here because it remains applicable even when some ODMR resonances are partially resolved or inaccessible within the experimental bandwidth. In this sense, the fitting procedure serves as an enabling tool for extracting stray-field variations associated with domain reversal under realistic experimental constraints.

We begin with the effective ground-state Hamiltonian of a single NV center
\begin{equation}
    \mathcal{H} = D S_{z}^{2} + E (S_{x}^{2} - S_{y}^{2}) + \gamma_{e} \vec{B} \cdot \vec{S},
\label{Hg}
\end{equation}
where $S$ is the spin-1 operator, $D$ and $E$ are the zero-field splittings respectively denoting the center and the splitting of the energy levels in the absence of a magnetic field, $\gamma_{e} = 28$~MHz/mT is the electron gyromagnetic ratio, and $\vec{B}$ is the local magnetic field. Solving this Hamiltonian yields two transitions $\ket{0} \leftrightarrow \ket{\pm1}$. For each ND, the ODMR response is treated as the summed contribution of the four NV orientations. Because the NDs are randomly oriented, the absolute orientation of the crystal axes in the laboratory frame is not required explicitly. Instead, the problem is formulated in a local NV frame, with the only external constraint being that the applied magnetic field has a known direction in the laboratory frame, namely perpendicular to the sample plane.

In the full model, the ODMR spectrum is described by ten parameters associated with two ZFS parameters, three magnetic field parameters, three Lorentzian lineshape parameters, and two effective contrast modulations. It is well established that the ODMR lineshape is strongly influenced by external driving and optical conditions. Accordingly, we model the combined effects of the static magnetic field, MW excitation, and laser illumination to reproduce the observed ODMR contrast. In practice, several of these can be calibrated independently and then held fixed across the full data set. Specifically, the ZFS parameters and intrinsic linewidth are obtained from spectra measured at a local zero-field condition, while two effective modulation parameters are determined from a high-signal, nondegenerate ODMR spectrum. After these calibrations, only five free parameters remain in the fit: the magnetic field magnitude $\norm{\vec{B}}$, polar angle $\theta_{B}$, azimuthal angle $\varphi_{B}$, baseline $y_{0}$, and contrast $C$. 

A central advantage of this procedure is that it uses the full ODMR spectrum, including partially overlapping or weak resonances, rather than relying on a subset of clearly resolved peak positions. This is especially important in the present measurements. In the low-field regime, several resonances become nearly degenerate, so the ODMR spectrum collapses toward only two broad resonances. In the high-field regime, some resonances shift outside the measurement window. Under both conditions, conventional multi-peak fitting becomes less reliable or requires additional assumptions. By contrast, the constrained fit remains relatively stable because the allowed spectral response is restricted by the Hamiltonian and by the calibrated lineshape model. As a result, the local magnetic-field magnitude can still be extracted robustly even when the number of visually resolved resonances is limited. (see Supporting Information for modeling details and justifications).

\subsection{Expected and unexpected stray magnetic field behaviors}
\begin{figure}[t]
    \includegraphics[width=12cm]{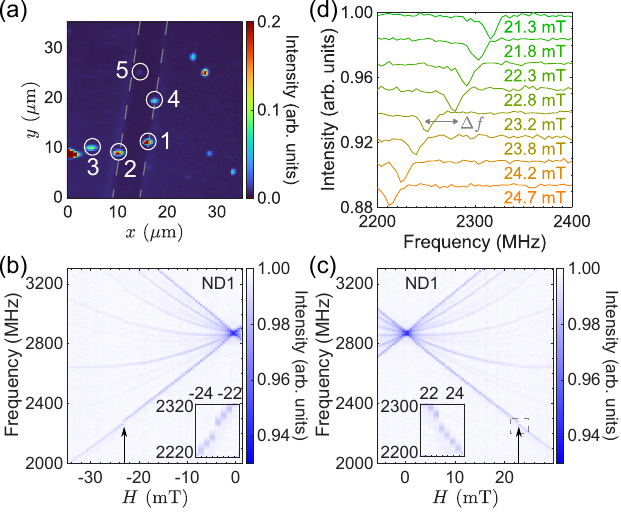}
    \caption{Expected frequency jumps with domain flips. (a) Confocal image showing NDs distributed on the CoFeB strip outlined by dashed-gray lines. Bright spots correspond to NDs; five are selected for subsequent measurements. ODMR color plots of ND1 during (b) negative and (c) positive magnetic field sweep. In both sweep directions, a clear NV frequency jump is observed, corresponding to a coercive field $H_{c} = 22.9\pm0.3$~mT. Insets: zoomed-in views of the frequency jump. The color scale represents photoluminescence intensity. (d) ODMR line plot of the dashed-gray region in (c), highlighting a sharp frequency increase $\Delta f$. The applied magnetic field $H$ is calibrated by ND3, which is not affected by the CoFeB sample.}
    \label{fig:fig2}
\end{figure}

A confocal image is shown in \cref{fig:fig2}(a), where the CoFeB strip is indicated by dashed-gray lines. Five NDs (ND1 - ND5) at different locations are investigated (see Supporting Information for the data of ND3 and ND4). As an initialization step, we first saturate the CoFeB device under a large external magnetic field $\abs*{H} > H_{c}$ and perform ODMR measurements during a gradual sweep of the applied magnetic field to the opposite polarity $\abs*{-H} > H_{c}$. Intuitively, a domain flip is expected to produce an increase in the local magnetic field as the magnetization aligns with the external field. Yet, this behavior is only observed in some NDs not all NDs.

\cref{fig:fig2}(b) and \cref{fig:fig2}(c) display ODMR color plots of negative and positive magnetic field sweeps of ND1, located at the edge of the CoFeB strip. A clear NV resonance frequency jump is observed near the coercive field of $H_{c}=22.9\pm0.3$~mT, in agreement with the independent MOKE measurement. To highlight this feature, ODMR line plots are presented in \cref{fig:fig2}(d), showing a sudden frequency increase $\Delta f$ near $H_{c}$. Using our fully constrained NV fitting procedure, the corresponding magnetic field change is estimated to be $\approx0.6$~mT.

\begin{figure}[t]
    \includegraphics[width=12cm]{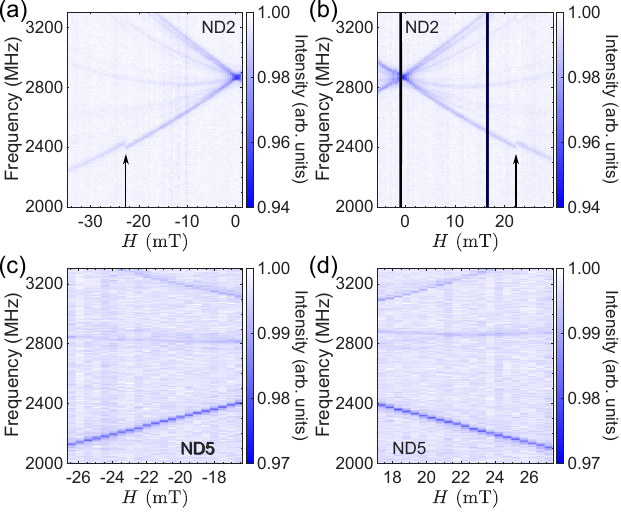}
    \caption{Spatially inhomogeneous stray fields measured by ND sensors. ODMR color plot of ND2 during (a) positive and (b) negative magnetic field sweep. In contrast to ND1, the observed frequency jump corresponds to a decrease in the local magnetic field upon domain flipping. Black vertical strips masked the missing data due to computer acquisition errors. (c, d) ODMR measurements of ND5 under (c) positive and (d) negative field sweeps, showing no discernible frequency jumps upon domain flipping. The color scale represents photoluminescence intensity.}
    \label{fig:fig3}
\end{figure}

We next examine the ODMR color plots of ND2 in \cref{fig:fig3}(a) and \cref{fig:fig3}(b). Although ND2, located just outside the CoFeB strip, again exhibits a clear frequency jump, the shift is opposite in sign, indicating a decrease in frequency (magnetic field) upon domain reversal. Furthermore, for ND5 located near the center of the CoFeB strip, no discernible magnetic field change is observed as shown in \cref{fig:fig3}(c) and \cref{fig:fig3}(d). This apparent ``transparency" of ND5 to a domain flip is counterintuitive. Based on these observations, we conclude that the stray magnetic field is highly position-dependent and the expected increase in frequency is not universally observed. We note that such pronounced spatial variation has not been reported in previous ferromagnetic resonance studies~\cite{Rable2022Local}. In the following, we show that these behaviors can be understood from the complex spatial profile of stray fields superimposed on the applied magnetic field.

\subsection{Elucidating the stray field distribution}
A CoFeB film with perpendicular magnetic anisotropy can be viewed as the magnetic analogue of a planar capacitor: just as the stray electric field of a planar capacitor is concentrated near its edge, the stray magnetic field of the film is primarily generated at its boundaries. For a semi-infinite thin film where the strip width $2w$ far exceeds the thickness $t$, i.e., $2w \gg t$, the stray magnetic field is given by~\cite{Hingant2015Measuring}
\begin{align}
    B_{x}(x, z) &= \hphantom{-}\frac{\mu_{0}M_{s}t}{2\pi}\frac{z}{(\abs{x}-w)^{2}+z^{2}},\nonumber\\
    B_{z}(x, z) &= -\frac{\mu_{0}M_{s}t}{2\pi}\frac{x}{(\abs{x}-w)^{2}+z^{2}},
\end{align}
where the edge is placed at $x = \pm w$, $z$ is the standoff distance (i.e NV-to-sample distance) perpendicular to the strip, $M_{s}$ is the saturation magnetization, and $\mu_{0}$ is the vacuum permeability. These expressions explicitly show that the in-plane field ($B_{x}$) is maximal at the sample edge, while the out-of-plane field ($B_{z}$) reaches its extrema at lateral positions $x = \pm w\pm z$. Both components vanish rapidly away from the edge, consistent with our experimental observations.

\begin{figure}[t]
    \includegraphics[width=12cm]{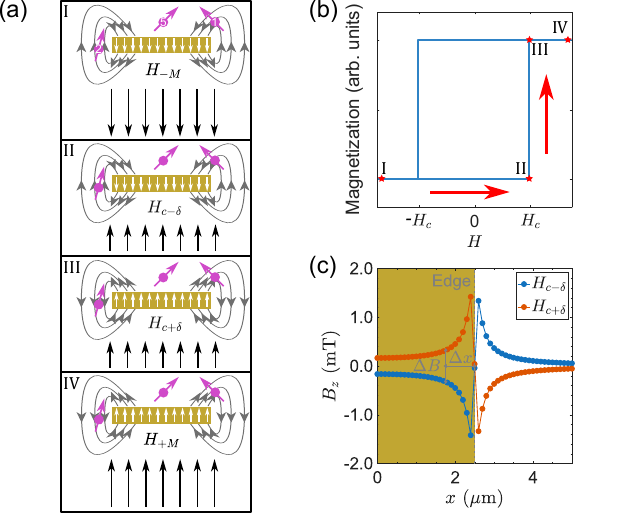}
    \caption{Origin of spatially dependent response from different ND sensors. (a) Schematic of local stray field (gray curves/arrows) upon domain reversal in a thin film with perpendicular magnetic anisotropy under an out-of-plane applied magnetic field (black arrows). The local field measured by a ND sensor is highly position dependent owing to the alignment of external and stray fields: negligible at the center (ND5), positive at the edge (ND1), and opposite to the applied field just outside the strip (ND2) during switching near the coercive field $H_{c}$. Labels (I, II, III, and IV) correspond to the magnetization states defined in (b). (b) Illustration of magnetization as a function of external $H$ field. Red arrows indicate the magnetic field sweep direction. (c) Micromagnetic simulations of the stray magnetic field before ($H_{c-\delta}$) and after ($H_{c+\delta}$) domain reversal at a standoff distance of $z=70$~nm (half of the averaged ND size). By matching the simulated field change to the measured value for ND1 ($\Delta B \approx 0.6$~mT), a most probable lateral distance of $\Delta x \approx 740$~nm from the sample boundary is obtained. $z$-axis is out of plane and $x$-axis is along the width of the CoFeB.}
    \label{fig:fig4}
\end{figure}

Different responses from multiple ND sensors can be further understood by considering the evolution of stray field lines under an sweeping external field, as sketched in \cref{fig:fig4}(a) with four representative stages labeled in \cref{fig:fig4}(b). Stage I and IV correspond to negatively and positively saturated magnetization, while stage II and III correspond to the external field just before and after the domain flip near the coercive field $H_{c}$. Starting from negative saturation $\abs*{H_{-M}} > H_{c}$ (I) and sweeping to positive fields $H_{+M} > H_{c}$ (IV), the stray magnetic field associated with the domain structure reverses its direction across the coercive field, i.e., between $H_{c-\delta}$ (II) and $H_{c+\delta}$ (III). This inversion produces a spatially dependent local magnetic field detectable by nearby NDs. For a film with $w \gg t$, NDs near the center experience negligible change due to the vanishing stray field, whereas NDs near the sample boundary detect strong signals: a positive jump at the edge and a negative jump just outside the strip.

To corroborate this conceptual explanation, we perform micromagnetic simulations in Object-Oriented Micro-Magnetic Framework (OOMMF) with a unit cell of $2 \times 2 \times 1$~nm$^{3}$. The magnetization dynamics are obtained by solving the Landau-Lifshitz-Gilbert equation. The material parameters for CoFeB are chosen as follows: saturation magnetization $M_{s}=1.3\times10^{6}$~A/m, damping constant $\alpha=0.008$, and exchange stiffness $A=1.8\times10^{-11}$~J/m. The sample is centered at $x = 0~\mu$m, and the edge is located at $x = 2.5~\mu$m. Various standoff distances and domain configurations were simulated (see Supporting Information for extended simulation data).

We assume a representative standoff distance of $z=70$~nm (half of the averaged ND size) and calculate the sample's stray magnetic field before ($H_{c-\delta}$) and after ($H_{c+\delta}$) the domain reversal. Both curves are plotted on the same graph in \cref{fig:fig4}(c) to show the stray magnetic field jump. As we discussed above, the $B_{z}$ component reaches its extrema slightly off from the edge. Moreover, the jump in the simulated fields during domain reversal can be used to estimate the position of the ND. For ND1 with $\Delta B \approx 0.6$~mT, the most probable lateral distance from the sample edge is $\Delta x \approx 740$~nm. For ND2, we notice that the frequency jump $>1~\rm{mT}$ is higher than ND1, which could be explained by a closer distance $\sim300$~nm from the edge. It is worth mentioning that, due to the strong gradient fields, it is critical to have a small ND for probing the local magnetic field. The choice of 140-nm ND is suitable for capturing such a spatially varying stray field.

\subsection{Comparison with other NV techniques}
In general, NV-SPM provides exceptional spatial resolution and sensitivity for investigating nanoscale magnetism and validating microscopic models~\cite{Rondin2013Stray, Tetienne2015The, Gross2016Direct, Jenkins2019Single, Tsukamoto2025Observation, Thiel2019Probing, Dovzhenko2018Magnetostatic, Hingant2015Measuring, Wu2025Nanoscale}, whereas NV-WFM offers a large field-of-view with diffraction-limited spatial resolution for imaging spatial inhomogeneities~\cite{Dailledouze2025Imaging, Schlussel2018Wide, Lillie2020Laser, Broadway2020Imaging, Healey2022Varied, Huang2023Revealing, McLaughlin2021Strong, McLaughlin2022Quantum, Chen2025Widefield}. Positioned between these two approaches, ND-based sensing combines nanoscale proximity with experimental simplicity. To date, ND-based studies have focused mainly on domain wall oscillations~\cite{Rable2023Off}, ferromagnetic resonances~\cite{Rable2022Local, Page2019Optically, Wolfe2014Off}, and spin-waves~\cite{Wu2024Wideband, Andrich2017Long, Du2017Contorl, Lee-Wong2020Nanoscale}.

Compared to NV-SPM and NV-WFM, ND-based sensing offers several practical advantages. NDs can be deposited directly onto devices and transferred readily between experimental setups, without the specialized handling required for bulk diamond probes. By contrast, NV-embedded diamond tips and plates are costly, fragile, and often require system-specific calibration. These features make NDs a simple, low-cost, and adaptable platform for magnetic device characterization. ND-based sensing also provides a favorable route to high spatial resolution: unlike NV-WFM, whose resolution is fundamentally limited by optical diffraction ($\approx 500$~nm), the resolution of ND-based measurements is set primarily by the ND size. Commercial NDs with diameters of only a few tens of nanometers therefore enable spatial resolution approaching that of NV-SPM. Although NV-SPM can access similar length scales, its performance is often affected by tip-to-tip variability, including differences in NV properties, stand-off distance, and tip geometry, which require careful calibration~\cite{Xu2025Minimizing}. ND-based sensing likewise benefits from screening for suitable particles, and high-quality NDs can in principle support pulsed protocols for improved sensitivity. Taken together, these attributes position ND-based sensing as a practical and scalable alternative for quantitative nanoscale magnetometry. This advantage is especially important for domain-wall devices, where magnetic strips can be narrowed to hundreds or even tens of nanometers. In this regime, NV-WFM is fundamentally limited by optical diffraction ($\approx 500$~nm), while three-dimensional device geometries may complicate the use of NV-SPM. ND-based magnetometry is therefore particularly well suited to such structures because of its simplicity and the availability of commercial NDs as small as 30~nm.

\section{Conclusions}
In conclusion, we demonstrate ND-based sensing on a CoFeB magnetic device and use 140-nm NDs at different locations to resolve strongly position-dependent stray fields during magnetic domain reversal. We identify three distinct responses: a positive magnetic-field jump at the sample edge, an opposite-sign negative jump just outside the strip, and no measurable change near the strip center. These observations are quantitatively reproduced by independent micromagnetic simulations, validating both the physical interpretation and the robustness of the sensing approach. Our results establish ND-based magnetometry as a powerful platform for quantitative, spatially resolved characterization of heterogeneous magnetic fields in patterned magnetic devices.

More broadly, this proof-of-principle work identifies ND-based sensing as a promising diagnostic for emerging magnetic device concepts, particularly domain-based neuromorphic architectures. In these systems, short-range and spatially inhomogeneous stray fields are difficult to distinguish from electrically generated fields, yet that distinction is essential for minimizing crosstalk and enabling scalable operation. By directly resolving such local field variations during switching, ND-based sensing provides a practical route to understanding, engineering, and optimizing the magnetic field landscape at the device level. In this sense, the method is not only a characterization tool, but also an enabling capability for the rational design of next-generation magnetic technologies.

\section{Methods}
\subsection{Sample stack}
The sample stack consists of substrate/\ce{SiO_{2}}(100~nm)/Ta(10~nm)/\ce{Co_{20}Fe_{60}B_{20}}(1.2~nm)/MgO($\sim1$~nm). The interfacial coupling turns the CoFeB strip into perpendicular magnetic anisotropy. The CoFeB strip has approximate dimensions of 5~$\mu$m in width and 83~$\mu$m in length. Notches are introduced to facilitate domain wall nucleation. An external static magnetic field is applied perpendicular to the film plane and swept between approximately -35 mT and +30 mT, depending on the sweep direction.

\subsection{NV measurements}
The laser power is maintained at $\approx11~\mu$W throughout measurements. MW excitation for NV spin manipulation is delivered via a Pt wire with a diameter of $\sim70~\mu$m. The MW power is kept at +24~dBm. The lateral distance between the center of the wire and the center of the CoFeB strip is $\sim80~\mu$m. NDs (Adamas Nanotechnologies, Inc., NDNV140nmHi) are drop-cast onto the CoFeB device from a solution with a density of $10~\mu$g/mL. The sensitivity of the NDs~\cite{Rondin2014Magnetometry} around the coercive field at $\approx 20$~mT is $\approx20~\mu\mathrm{T/\sqrt{Hz}}$. Throughout the main text, the applied magnetic field is calibrated using ND3, an ND far from the CoFeB strip which experiences a negligible sample stray magnetic field.

\section{Author contributions}
J.A.C.I. and X.L. conceived and designed the experiments. K.O.H. performed all measurements and analyzed the data. H.-P.C. contributed to the MOKE measurement. C.C. provided the CoFeB samples. H.A. and A.P. contributed to preliminary measurements. L.J.C. contributed to the micromagnetic simulations. J.T. contributed to the explanation of stray magnetic fields. K.O.H. and X.L. wrote the manuscript with assistance from all other authors.

\begin{acknowledgement}
The work performed by K.O.H and X.L. has been primarily supported by the Air Force Office of Scientific Research (AFOSR) under MURI grant number FA9550-25-1-0288. H.A. and A.P. have been primarily supported by the NSF Designing Materials to Revolutionize and Engineer our Future (DMREF) program via grants DMR-2118806 (building experimental setup). J.T. is supported by funding from the Center for Energy Efficient Magnonics an Energy Frontier Research Center funded by the U.S. Department of Energy, Office of Science, Basic Energy Sciences at SLAC National Laboratory under contract DE-AC02-76SF00515 (dynamics in magnetic materials). X.L. gratefully acknowledges funding from the Welch Foundation Chair F-0014 (materials supply). H.-P.C., C.C., and J.A.C.I. acknowledge funding from the US National Science Foundation (NSF) under Award ECCS-2343606, and fabrication resources from the UT Austin Microelectronics Research Center (MRC), partially funded by the NSF National Nanotechnology Coordinated Infrastructure (NNCI) via Grant ECCS 2025227.

\end{acknowledgement}

\begin{suppinfo}
The Supporting Information is available free of charge.

Description of the fully constrained NV fitting, extended data for ND3 and ND4, and extended data for micromagnetic simulations.

\end{suppinfo}

\bibliography{references}
\newpage

\end{document}